\documentclass[preprint,10pt]{elsarticle}

\graphicspath{{./Figures/}}

\biboptions{comma,sort&compress}

\journal{Technical Report ZJU (undergoing update)}

\begin{document}

\begin{frontmatter}

\title{Parametric/direct CAD integration}

\author[]{Qiang Zou}
\ead{john.qiangzou@gmail.com}

\address[]{State Key Laboratory of CAD$\&$CG, Zhejiang University, Hangzhou, 310027, China}

\begin{abstract}
In the history of computer-aided design (CAD), feature-based parametric modeling and boundary representation-based direct modeling are two of the most important CAD paradigms, developed respectively in the late 1980s and the late 2000s. They have complementary advantages and limitations, thereby offering huge potential for improvement towards an integrated CAD modeling scheme. Some believe that their integration will be the key characteristic of next generation CAD software. This paper provides a brief review on current parametric/direct integration approaches. Their basic ideas, advantages, and disadvantages will be discussed. The main result reads that existing integration approaches are far from being completed if seamless parametric/direct integration is desired. It is hoped that, by outlining what has already been made possible and what still remains problematic, more researchers will be attracted to work on this very important research topic of parametric/direct integration. 

This paper serves as a complement to the CAD paper titled ``Variational Direct Modeling: A Framework Towards Integration of Parametric Modeling and Direct Modeling in CAD." Please cite this work as follows:
Qiang Zou, Hsi-Yung Feng, and Shuming Gao. Variational Direct Modeling: A Framework Towards Integration of Parametric Modeling and Direct Modeling in CAD. Computer-Aided Design 157 (2023): 103465.

\end{abstract}


\end{frontmatter}

\section{Introduction}
\label{sec:introduction}
A product is a physical artifact; a product model is a computerized representation of a product. Computer-aided design (CAD) is the use of computers to construct and edit product models. CAD has been the dominant tool used in industry to support engineering practices like modeling \cite{dieter2009engineering}, analysis \cite{cottrell2009isogeometric}, optimization \cite{rao2019engineering}, manufacturing \cite{zou2014iso}, maintenance \cite{dhillon2002engineering}, inspection \cite{zou2021robusttool} and design/processes reuse \cite{denkena2007knowledge,zou2013iso,su2020accurate,zou2022variational}.  The notion of CAD may date back to the 1960s \cite{coons1963outline, ross1960computer,sutherland1964sketchpad}, which arose out of (1) a need to digitally describe products for downstream applications, e.g., numeric control machine tools \cite{shah1995parametric} and (2) a goal to develop a human-computer cooperation system based on product models for solving engineering design problems \cite{coons1960computer}, which may be summarized as human-machine communication.

CAD modeling, in its broadest sense, is concerned with modeling all aspects of a product's information needed to support processes throughout the product's life cycle. These aspects consist primarily of the product's spatial information, augmented with secondary, non-spatial information such as material, hardness, and so forth. In a narrow sense, CAD modeling focuses merely on modeling spatial information, which is sometimes called
geometric modeling. This definition of CAD modeling is more commonly used. To date, major CAD modeling schemes may be classified into five categories: wireframe modeling, surface modeling, solid modeling, parametric modeling, and direct modeling.

Wireframe modeling, surface modeling, and solid modeling may be seen as the preliminary stage in CAD development. They focus on the mathematical representation of a product's spatial information. Solid modeling as the last and the most important representation scheme has the distinctive property of representational completeness, which can contribute to automatically answering any possible geometric queries \cite{Shapiro2002}. It began in the 1970s with two competing approaches \cite{hoffmann2005constraint}: constructive solid geometry (CSG) and boundary representation (B-rep). They have respectively motivated parametric modeling in the late 1980s and direct modeling in the late 2000s.

Parametric modeling combines associative modeling and solid modeling to achieve the goal of automatic change propagation. Associative modeling allows specification of associativity between geometric entities in a solid model. A locally made change (taking the form of parameter changes) can then propagate automatically through the associativity network. With automatic change propagation come the benefits of geometry reuse and embedding of engineering knowledge with geometry \cite{shah1995parametric}. There are, however, serious limitations of parametric modeling, including (1) managing modeling history requires considerable expertise, (2) updating models (i.e., regenerating modeling history) is often time-consuming, and (3) editing a model is difficult if the model was designed with an intention that conflicts with the desired changes \cite{camba2016parametric,ElHani2012,monedero2000parametric}.

Direct modeling, a very recent CAD modeling paradigm, is often presented as the opposite of parametric modeling. All direct modeling focuses on is to provide flexible model edits and fast model update. This is essentially achieved by discarding the use of associativity and making all geometric entities in a solid model free to change. Typically, users just need to grab, push, and pull the geometric entities of interest so as to make desired changes \cite{Zou,zou2021robust}. With this geometry-oriented strategy, users can directly edit the model as they see fit without considering how the model was constructed. This unprecedented model editing flexibility is, however, attained at a high price: without associativity, direct modeling barely supports parametric model edits.

Clearly, parametric and direct modeling have complementary advantages and limitations. Their integration can thus provide the best of both worlds \cite{zou2020decision,Fu2017}, and will be one of the key characteristics of the next generation CAD software. Of course, some other directions are also promising, e.g., CAD for 3d printed parts \cite{liu2021memory,ding2021stl,gupta2018quador}, generative design \cite{krish2011practical}, CAD+AI \cite{oh2019deep}, to name a few. To date, there have been five integration approaches reported by industry and academia. However, none of the approaches meets the need of seamless parametric/direct integration. In what follows, a brief introduction to the basic notions of solid modeling, parametric modeling and direct modeling, and a review on parametric/direct integration are to be presented.

\section{A brief on solid modeling}
\label{sec:solid-modeling}
As already noted, spatial information has always played a fundamental role in mechanical design; but what exactly does spatial information mean in the domain of mechanical design? It is impossible to have effective CAD modelers unless the answer to this question is made precise. Solid modeling emerged as a response to this need, which was established by Requicha and Voelcker, then at University of Rochester, and Braid and Lang, then in Cambridge \cite{braid1975synthesis,voelcker1977geometric}.

\begin{figure}[t]
  \centering
  \hfill
  \includegraphics[width=\textwidth]{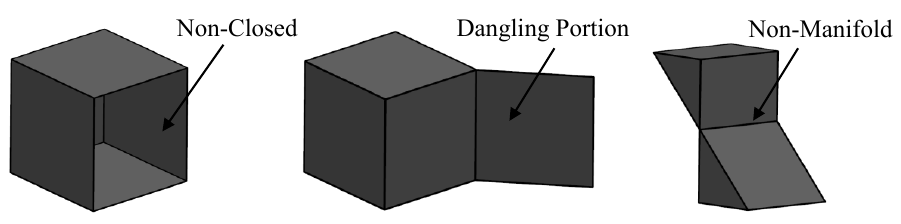}
  \hspace*{\fill}
  \caption{Illustrations of the notions of closed (left), regular (middle), and manifold (right).}\label{fig:solid-notions}
\end{figure}

A solid is a physical object occupying certain volume in 3D Euclidean space $R^3$. A volume is a subset of $R^3$, but not all subsets of $R^3$ are volumes. A subset has to satisfy certain conditions that encapsulate the characteristics of ``solidity" to be deemed a volume. A widely accepted set of conditions are given in \cite{Requicha1980}: a solid is a subset of $R^3$ that is bounded, closed, regular, and semi-analytic, often referred to as a r-set. Let the subset be represented by $s\subset R^3$. The ``bounded'' condition means that $s$ must occupy a finite portion of space; the other three conditions require that the boundary between $s$ and $R^3 - s$ is well-behaved, where the operator $-$ is set difference. To be more precise, the ``closed'' condition means that $s$ contains its boundary, see Fig.~\ref{fig:solid-notions} for a non-closed example. The ``regular'' condition means that the boundary of $s$ does not contain any dangling portions as shown in Fig.~\ref{fig:solid-notions}; mathematically, ``regular'' means that $s$ equals to the closure of its interior. The ``semi-analytic'' condition means that the boundary of $s$ is composed of piecewise analytic surfaces (which are surfaces that can be described by analytic functions). These conditions encompass the common attributes of most objects in mechanical design \cite{Requicha1977}. However, they allow non-manifold models, which are not manufacturable \cite{Mantyla1984a}. For this reason, many prefer a solid to be a r-set with a manifold boundary. A boundary is deemed manifold if, for each point on the boundary, there is an open neighborhood that can be continuously deformed to a 2D disk (mathematically, homogeneous to an open disk in $R^2$). Fig.~\ref{fig:solid-notions} shows a counterexample to manifold boundary.

\begin{figure}[t]
  \centering
  \hfill
  \includegraphics[width=\textwidth]{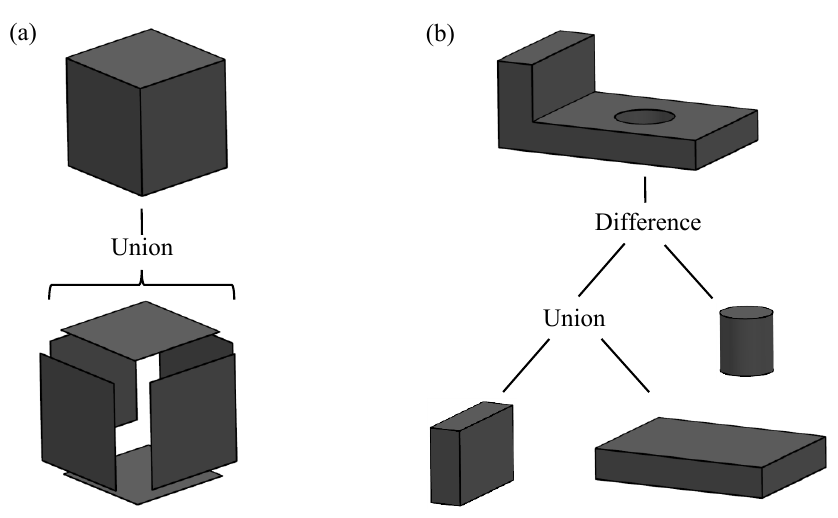}
  \hspace*{\fill}
  \caption{Examples of B-rep (a) and CSG (b).}\label{fig:brepcsg}
\end{figure}

A solid model is a computerized representation of a solid. To date, several representation schemes for solids have been made available, as reviewed in \cite{Requicha1980}. Among them, the mainstream schemes are boundary representation (B-rep) and constructive solid geometry (CSG). B-rep, as the name indicates, describes a solid using the boundary between the solid and non-solid. A B-rep solid model is essentially a collection of interconnected faces comprising the solid's boundary (Fig.~\ref{fig:brepcsg}a). There exist various schemes for representing boundary faces. The basic scheme stores the boundary faces' carrying surfaces and connections between them \cite{Braid1986}. Connections define how a carrying surface is intersected with other surfaces and trimmed to a bounded surface. Usually, we refer to the carrying surfaces as geometry and the connections as topology. Hereafter, surfaces default to carrying surfaces such as infinite planes; faces default to bounded surfaces. Alternative schemes have also been proposed for the purpose of speeding up specific algorithms \cite{Requicha1980}. They differ from the basic scheme in the redundant information stored. Typical redundant information takes the form of faces' bounding edges and vertices, as well as connections between them.

CSG represents a solid as successive combinations of primitive shapes via (regularized) Boolean operations  \cite{Requicha1985}, as shown in Fig.~\ref{fig:brepcsg}b. CSG is often implemented as binary trees where leaves represent primitive shapes, and nodes Boolean operations. Different from B-rep that stores the solid's boundary explicitly, CSG is an implicit representation scheme: it does not store the solid's final spatial information but the construction history. Due to the use of Boolean operations, CSG guarantees model validity \cite{Requicha1985}; by contrast, B-rep does not have guaranteed validity.

\section{A brief on parametric and direct modeling}
\label{sec:parametric-direct-modeling}
When solid modeling-based CAD systems were introduced, it met with resistance from users due to the difficulty of use. It was not until the introduction of parametric modeling-based CAD systems that this resistance began to melt away \cite{shah1998designing}. The primary benefit of parametric modeling is to allow users to attain solid model variants through editing parameters embedded in the model \cite{camba2016parametric}. The embedding is done by constructing associativity between geometric entities in the model; the associativity often takes the form of unidirectional and/or bi-directional geometric constraints. With the associativity, a parameter change will propagate automatically. Also, the user can design the propagation, the so-called design intent (user-defined admissible modifications of the model geometry). This shifted CAD from an instance modeler to an ``electronic master" modeler.

The mainstream implementation approach of parametric modeling involves three major procedures \cite{shah1998designing,camba2016parametric}: (1) 2D sketching, (2) 3D sweeping, and (3) feature combining. The user defines a planar topology and specifies 2D bi-directional geometric constraints relating geometric entities on the 2D sketch. The 2D sketches are then used as sections for sweeping to be 3D solids, which are called features in most modelers. These features are similar to solid components in CSG. A new 2D sketch plane will be positioned relatively to features already specified by the user via 3D unidirectional geometric constraints. A newly generated feature is combined with preceding features through Boolean operations, which is again similar to CSG. The sequence of performing the above three steps by the user is conventionally referred to as model construction history. When a change is made, the 2D sketches are re-evaluated, then the construction history is replayed, and finally a new solid model is generated.

The construction history will restrict the user to what can be edited and will determine how changes are propagated, thereby defining a model variation space \cite{shah1998designing}. Flexibility can be said to model edits falling in the space as a simple parameter change can do the job, but cannot be said to those outside the space. Thus, parametric modeling requires careful preplanning about the space. However, many model edits are likely hard to plan in advance, especially for fine model tuning. Making such unplanned model edits would require structural alterations to the construction history, which is very difficult and time-consuming if there is no good understanding of the modeling principles parametric modeling follows and the model's construction history \cite{camba2016parametric}. Updating it is time-consuming due to the complete model regeneration mechanism. All of these could lead to serious productivity issues in scenarios like collaborative design and editing others' designs \cite{camba2016parametric,ElHani2012,monedero2000parametric}.

The very recent direct modeling paradigm was proposed to achieve flexible model editing. ``Direct'' comes from this modeling scheme's main feature: interacting directly with the geometry of the model to make edits. Although initially introduced by industry, the direct modeling notion may be traced back to the local operations developed by the academic community in the 1980s \cite{Grayer1980,Rossignac1990,Stroud2006}. Some widely used local operation examples in today's CAD systems are filleting and chamfering. The very local operation that direct modeling stemmed from is the tweaking operation, which is the predecessor of push-pull operations. The essential advance made is: tweaking does not allow any violations to the pre-edit model topology, while push-pulling allows such violations. With this relaxation, direct modeling achieves much improved modeling flexibility, refer to \cite{Zou,zou2021robust} for examples. The early version of direct modeling consisted primarily of push-pull operations. The recent version includes more operations, most of which are just specialized push-pull operations such as the make-face-parallel operation. Despite its powerfulness, direct modeling lacks the capability of parametric edits \cite{zou2019variational,zou2020decision}.

\section{Parametric/direct integration}
\label{sec:parametric-direct-integration}
Some efforts, mostly from industry, have been made to allow users to have both direct and parametric capabilities in a same modeler. To the best of the authors' knowledge, there are five publicly reported integration approaches, as summarized below. 

\subsection{Pseudo-Features}
\begin{figure}[htbp]
  \centering
  \hfill
  \includegraphics[width=\textwidth]{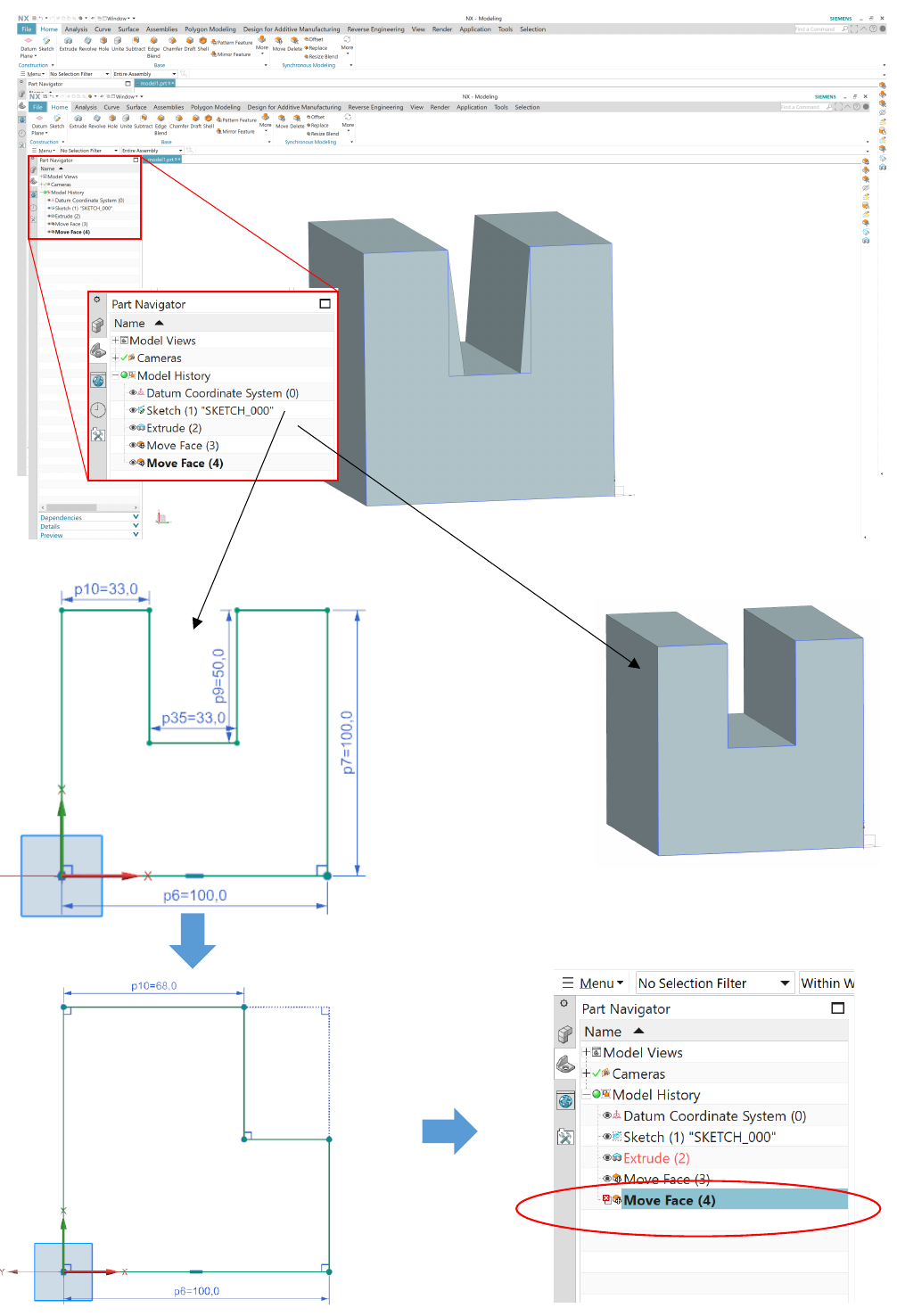}
  \hspace*{\fill}
  \caption{Illustration of the pseudo-feature method and its limitations.}\label{fig:pseudo-feature}
\end{figure}

This approach, proposed by industry, allows the user to simulate direct-modeling within a history-based parametric modeler. The user can modify models utilizing direct modeling methods (e.g., pushing and pulling faces) or parameter modeling methods. User-specified direct edits will be added to the end of the model's construction history as pseudo-features, and the original history remains exactly as before. Fig.~\ref{fig:pseudo-feature} shows an example of modeling a slot in this way (using Siemens NX). The zoomed-in view of the model's history shows clearly that the two Move-Face direct edits are just appended to the end of the history. This approach has been adopted by most CAD vendors. The primary advantage of this approach is that it is very easy to implement.

However, simply adding direct edits to the end of the model's history creates complex modeling history with questionable value to the CAD model. This could mess up the model's parametric information and lead to loss of meaningful parametric controls. Fig.~\ref{fig:pseudo-feature} shows an example of this situation. Changing the P10 dimension from 33mm to 68mm leads to a failed history regeneration, as shown in the bottom-right subfigure. The perfect solution to this problem is not adding direct edits to the end of model history, but transforming them into appropriate redefinition of relevant features. For this example, we should modify the 2D sketch, from a rectangular slot to a slanted slot.

\subsection{Dual Modes}
This approach, proposed by industry as well, allows users to switch between direct and parametric modeling modes in a same modeler. When in the direct modeling mode, the user can modify geometries utilizing direct modeling techniques (e.g., pushing and pulling faces). When in the parametric modeling mode, the user can modify parameters using feature modeling techniques. Its implementation is very simple: when switching from the parametric mode to the direct mode, the parametric model is downgraded to a dumb B-rep model; then direct modeling becomes applicable. However, it cannot recover any parametrics when switching back from the direct mode to the parametric mode \cite{nag2015methods}.

\subsection{Synchronous Technology}
This approach, again, proposed by industry (or more specifically, Siemens NX) can be viewed as an improvement to the above method. Instead of converting the whole parametric model to a B-rep model, it does a partial conversion. Features are separated into two types: direct-edit features and ordinary features. Only direct-edit features are converted to a B-rep model. Also, this approach places the direct-edit features before the ordinary features in the model history. The user can move an ordinary feature to the direct-edit feature set so to carry out direct modeling operations, but when one of those features is moved from the ordinary feature set to the direct-edit feature set, all ordinary features created prior to the feature being moved must also be moved (done automatically in the background)\cite{Chad2008}. Clearly, this causes unnecessary loss of meaningful parametrics.

It should be noted that there was a time when synchronous technology also referred to 3D variational modeling \cite{lin1981variational,chung2000framework}, but this interpretation appears not have been implemented in the current Siemens NX software package.

\subsection{Operation Translating} 
This approach, proposed by both academia and industry (Autodesk), translates direct edits into operations of parameter tuning and/or order rearrangement of the features already presented in the model history \cite{Fu2017,qin2021automatic,ushakov2008variational}. This way of working may help but cannot solve the problem altogether because not all direct edits are achievable through those feature operations. To make matters worse, feature parameter tuning for a given direct edit (if achievable) is usually not unique \cite{zou2020decision}.

\subsection{Constrained Direct Modeling}
This approach allows users to apply direct edits while keeping all geometric constraints of the model in the background, through efficient geometric constraint solving \cite{ushakov2008variational}. The geometric constraints are applied directly to the boundary elements of a solid model (faces, edges, vertices). And those constraints are generated by automatic constraint recognition algorithms. The disadvantage of this method is that the recognized geometric constraints usually differ from the original design intent. Also, solving 3D geometric constraint systems remains an open issue in its own right \cite{tang2022review,zou2019limitations}. Even worse, this method in its essence downgrades direct modeling to merely a graphical user interface (GUI) tool for making parametric modifications.

\section{Conclusion}
\label{sec:conclusion}
It can be concluded from the above review that, at the level of individual modeling techniques, substantial progress has been made for both parametric and direct modeling, but at the integration level, no satisfactory solution has been made available now if seamless parametric/direct integration is desired. Overall, current approaches/techniques attempting to address the integration problem are still at the early stage, and they all have inherent drawbacks.

Their failed attempts to seamlessly integrate parametric and direct modeling is due to the focus on indirect, conversion-based and easy-to-implement integration strategies, rather than working on strategies that can directly solve the challenges of parametric/direct integration. The primary challenge of integrating parametric and direct modeling lies in maintaining information consistency in a CAD model undergoing parametric/direct edits \cite{qiang2019variational}. A CAD model consists of information on topology, geometry, and constraints. Parametric and direct edits work at different layers of information, i.e., the constraint layer and the geometry layer, respectively. Changes at a layer cannot be  reflected in others by current model representation schemes. As a result, information inconsistencies are generated, and unpredictable modeling behavior and invalid models, is caused.

\section*{Acknowledgements}
This work is a byproduct of the authors developing a new CAD modeling approach: variational direct modeling. The associated project was in part funded by the Natural Sciences and Engineering Research Council of Canada (NSERC). This financial support is greatly appreciated.

\bibliographystyle{elsarticle-num}
\bibliography{Bibliography}

\begin{thebibliography}{10}
\expandafter\ifx\csname url\endcsname\relax
  \def\url#1{\texttt{#1}}\fi
\expandafter\ifx\csname urlprefix\endcsname\relax\def\urlprefix{URL }\fi
\expandafter\ifx\csname href\endcsname\relax
  \def\href#1#2{#2} \def\path#1{#1}\fi

\bibitem{dieter2009engineering}
G.~E. Dieter, L.~C. Schmidt, et~al., Engineering design, Vol.~4, McGraw-Hill
  Higher Education Boston, 2009.

\bibitem{cottrell2009isogeometric}
J.~A. Cottrell, T.~J. Hughes, Y.~Bazilevs, Isogeometric analysis: toward
  integration of CAD and FEA, John Wiley \& Sons, 2009.

\bibitem{rao2019engineering}
S.~S. Rao, Engineering optimization: theory and practice, John Wiley \& Sons,
  2019.

\bibitem{zou2014iso}
Q.~Zou, J.~Zhang, B.~Deng, J.~Zhao, Iso-level tool path planning for free-form
  surfaces, Computer-Aided Design 53 (2014) 117--125.

\bibitem{dhillon2002engineering}
B.~S. Dhillon, Engineering maintenance: a modern approach, cRc press, 2002.

\bibitem{zou2021robusttool}
Q.~Zou, Robust and efficient tool path generation for machining low-quality
  triangular mesh surfaces, International Journal of Production Research
  59~(24) (2021) 7457--7467.

\bibitem{denkena2007knowledge}
B.~Denkena, M.~Shpitalni, P.~Kowalski, G.~Molcho, Y.~Zipori, Knowledge
  management in process planning, CIRP annals 56~(1) (2007) 175--180.

\bibitem{zou2013iso}
Q.~Zou, J.~Zhao, Iso-parametric tool-path planning for point clouds,
  Computer-Aided Design 45~(11) (2013) 1459--1468.

\bibitem{su2020accurate}
C.~Su, X.~Jiang, G.~Huo, Q.~Zou, Z.~Zheng, H.-Y. Feng, Accurate model
  construction of deformed aero-engine blades for remanufacturing, The
  International Journal of Advanced Manufacturing Technology 106~(7) (2020)
  3239--3251.

\bibitem{zou2022variational}
Q.~Zou, Q.~Zheng, Z.~Tang, S.~Gao, Variational design for a structural family
  of cad models, arXiv preprint arXiv:2201.02926 (2022).

\bibitem{coons1963outline}
S.~A. Coons, An outline of the requirements for a computer-aided design system,
  in: Proceedings of the May 21-23, 1963, spring joint computer conference,
  1963, pp. 299--304.

\bibitem{ross1960computer}
D.~T. Ross, Computer-aided design: A statement of objectives, MIT Electronic
  Systems Laboratory, 1960.

\bibitem{sutherland1964sketchpad}
I.~E. Sutherland, Sketchpad a man-machine graphical communication system,
  Simulation 2~(5) (1964) R--3.

\bibitem{shah1995parametric}
J.~J. Shah, M.~M{\"a}ntyl{\"a}, Parametric and feature-based CAD/CAM: concepts,
  techniques, and applications, John Wiley \& Sons, 1995.

\bibitem{coons1960computer}
S.~A. Coons, R.~W. Mann, Computer-aided design related to the engineering
  design process, MIT Electronic Systems Laboratory, 1960.

\bibitem{Shapiro2002}
V.~Shapiro, {Solid modeling}, in: Handbook of Computer Aided Geometric Design,
  North-Holland, 2002, pp. 473--518.

\bibitem{hoffmann2005constraint}
C.~M. Hoffmann, Constraint-based computer-aided design, Tech. rep., Purdue
  University (2005).

\bibitem{camba2016parametric}
J.~D. Camba, M.~Contero, P.~Company, Parametric cad modeling: an analysis of
  strategies for design reusability, Computer-Aided Design 74 (2016) 18--31.

\bibitem{ElHani2012}
M.~A. {El Hani}, L.~Rivest, R.~Maranzana, {Product data reuse in product
  development: a practitioner's perspective}, in: IFIP International Conference
  on Product Lifecycle Management, Springer, 2012, pp. 243--256.

\bibitem{monedero2000parametric}
J.~Monedero, Parametric design: a review and some experiences, Automation in
  Construction 9~(4) (2000) 369--377.

\bibitem{Zou}
Q.~Zou, H.-Y. Feng, {Push-pull direct modeling of solid CAD models}, Advances
  in Engineering Software 127 (2019) 59--69.

\bibitem{zou2021robust}
Q.~Zou, H.-Y. Feng, A robust direct modeling method for quadric b-rep models
  based on geometry--topology inconsistency tracking, Engineering with
  Computers (2021) 1--16.

\bibitem{zou2020decision}
Q.~Zou, H.-Y. Feng, A decision-support method for information inconsistency
  resolution in direct modeling of cad models, Advanced Engineering Informatics
  44 (2020) 101087.

\bibitem{Fu2017}
J.~Fu, X.~Chen, S.~Gao, {Automatic synchronization of a feature model with
  direct editing based on cellular model}, Computer-Aided Design and
  Applications 14~(5) (2017) 680--692.

\bibitem{liu2021memory}
S.~Liu, T.~Liu, Q.~Zou, W.~Wang, E.~L. Doubrovski, C.~C. Wang, Memory-efficient
  modeling and slicing of large-scale adaptive lattice structures, Journal of
  Computing and Information Science in Engineering 21~(6) (2021).

\bibitem{ding2021stl}
J.~Ding, Q.~Zou, S.~Qu, P.~Bartolo, X.~Song, C.~C. Wang, Stl-free design and
  manufacturing paradigm for high-precision powder bed fusion, CIRP Annals
  70~(1) (2021) 167--170.

\bibitem{gupta2018quador}
A.~Gupta, G.~Allen, J.~Rossignac, Quador: Quadric-of-revolution beams for
  lattices, Computer-Aided Design 102 (2018) 160--170.

\bibitem{krish2011practical}
S.~Krish, A practical generative design method, Computer-Aided Design 43~(1)
  (2011) 88--100.

\bibitem{oh2019deep}
S.~Oh, Y.~Jung, S.~Kim, I.~Lee, N.~Kang, Deep generative design: Integration of
  topology optimization and generative models, Journal of Mechanical Design
  141~(11) (2019).

\bibitem{braid1975synthesis}
I.~C. Braid, The synthesis of solids bounded by many faces, Communications of
  the ACM 18~(4) (1975) 209--216.

\bibitem{voelcker1977geometric}
H.~B. Voelcker, A.~A. Requicha, Geometric modeling of mechanical parts and
  processes, Computer 10~(12) (1977) 48--57.

\bibitem{Requicha1980}
A.~G. Requicha, Representations for rigid solids: Theory, methods, and systems,
  ACM Computing Surveys (CSUR) 12~(4) (1980) 437--464.

\bibitem{Requicha1977}
A.~Requicha, Mathematical models of rigid solid objects, Tech. rep., Rochester
  University (1977).

\bibitem{Mantyla1984a}
M.~Mantyla, {A note on the modeling space of Euler operators}, Computer Vision,
  Graphics, and Image Processing 26~(1) (1984) 45--60.

\bibitem{Braid1986}
I.~C. Braid, {Geometric modelling}, in: Advances in Computer Graphics I,
  Springer, 1986, pp. 325--362.

\bibitem{Requicha1985}
A.~A. Requicha, H.~B. Voelcker, Boolean operations in solid modeling: Boundary
  evaluation and merging algorithms, Proceedings of the IEEE 73~(1) (1985)
  30--44.

\bibitem{shah1998designing}
J.~J. Shah, Designing with parametric cad: classification and comparison of
  construction techniques, in: International Workshop on Geometric Modelling,
  Springer, 1998, pp. 53--68.

\bibitem{Grayer1980}
A.~R. Grayer, {Alternative approaches in geometric modelling}, Computer-Aided
  Design 12~(4) (1980) 189--192.

\bibitem{Rossignac1990}
J.~R. Rossignac, {Issues on feature-based editing and interrogation of solid
  models}, Computers and Graphics 14~(2) (1990) 149--172.

\bibitem{Stroud2006}
I.~Stroud, P.~C. Xirouchakis, {CAGD - Computer-aided gravestone design},
  Advances in Engineering Software 37~(5) (2006) 277--286.

\bibitem{zou2019variational}
Q.~Zou, H.-Y. Feng, Variational b-rep model analysis for direct modeling using
  geometric perturbation, Journal of Computational Design and Engineering 6~(4)
  (2019) 606--616.

\bibitem{nag2015methods}
A.~Nag, T.~D. Gallagher, J.~J. Dunne, Methods and systems for converting select
  features of a computer-aided design model to direct-edit features, uS Patent
  9,117,308 (2015).

\bibitem{Chad2008}
J.~Chad, H.~David, {Synchronous technology: the best of both worlds for
  engineering organizations}, Tech. rep., Aberdeen Group, Boston (2008).

\bibitem{lin1981variational}
V.~C. Lin, D.~C. Gossard, R.~A. Light, Variational geometry in computer-aided
  design, ACM SIGGRAPH 15~(3) (1981) 171--177.

\bibitem{chung2000framework}
J.~C. Chung, T.-S. Hwang, C.-T. Wu, Y.~Jiang, J.-Y. Wang, Y.~Bai, H.~Zou,
  Framework for integrated mechanical design automation, Computer-Aided Design
  32~(5-6) (2000) 355--365.

\bibitem{qin2021automatic}
X.~Qin, Z.~Tang, S.~Gao, Automatic update of feature model after direct
  modeling operation, Computer-Aided Design and Applications 18 (2021) 170--85.

\bibitem{ushakov2008variational}
D.~Ushakov, Variational direct modeling: how to keep design intent in history
  free cad, Tech. rep., LEDAS Ltd (2008).

\bibitem{tang2022review}
Z.~Tang, Q.~Zou, H.-Y. Feng, S.~Gao, C.~Zhou, Y.~Liu, A review on geometric
  constraint solving, arXiv preprint arXiv:2202.13795 (2022).

\bibitem{zou2019limitations}
Q.~Zou, H.-Y. Feng, On limitations of the witness configuration method for
  geometric constraint solving in cad modeling, arXiv preprint arXiv:1904.00526
  (2019).

\bibitem{qiang2019variational}
Q.~Zou, Variational direct modeling for computer-aided design, Ph.D. thesis,
  University of British Columbia (2019).

\end{thebibliography}

\end{document}